\documentclass[fleqn,twoside]{article}
\usepackage{espcrc2}

\usepackage{graphicx}
\usepackage[figuresright]{rotating}

\usepackage{cite}

\def\NPPS{Nucl. Phys. B (Proc. Suppl.)}

\def\MP{Int. J. Mod. Phys.}

\newcommand{\eqn}[1]{(\ref{#1})}
\newcommand{\be}{\begin{equation}}
\newcommand{\ee}{\end{equation}}
\newcommand{\no}{\nonumber}

\newcommand{\ba}{\begin{array}{c}}
\newcommand{\bat}{\begin{array}{cc}}
\newcommand{\ea}{\end{array}}
\newcommand{\beqn}{\begin{eqnarray}}
\newcommand{\eeqn}{\end{eqnarray}}

\newcommand{\bi}{\begin{itemize}}
\newcommand{\ei}{\end{itemize}}

\newcommand{\rms}{\rm\scriptsize}

\newcommand{\cO}{{\cal O}}

\newcommand{\cJ}{{\cal J}}

\newcommand{\AmS}{{\protect\the\textfont2
  A\kern-.1667em\lower.5ex\hbox{M}\kern-.125emS}}

\title{Recent Progress on Tau Lepton Physics}

\author{A. Pich\address{Departament de F\'{\i}sica Te\`orica, IFIC,
Univ. Val\`encia--CSIC, Apt. 22085, E-46071 Val\`encia, Spain}
}

\begin{document}

\begin{abstract}
Some important aspects of hadronic $\tau$ decays are reviewed:
the determination of $\alpha_s$ from the inclusive $\tau$ hadronic width,
the measurement of $|V_{us}|$ through the Cabibbo-suppressed decays
of the $\tau$, and the theoretical description of the $\tau\to\nu_\tau K \pi$ spectrum.
The present status of other relevant electroweak topics, such as charged-current universality tests
or bounds on lepton-flavour violation, has been already summarized in ref.~\cite{PhiPsi08}.
\vspace{1pc}
\end{abstract}

\maketitle

\section{The inclusive hadronic width of the tau}

The hadronic $\tau$ decays turn out to be a beautiful
laboratory for studying strong interaction effects at low energies \cite{taurev06,taurev98}.
The $\tau$ is the only known lepton massive enough to decay into hadrons.
Its semileptonic decays are then ideally suited for studying the
hadronic weak currents.

The inclusive character of the total $\tau$ hadronic width renders
possible an accurate calculation of the ratio
\cite{BR:88,NP:88,BNP:92,LDP:92a,QCD:94}
$$
 R_\tau \equiv { \Gamma [\tau^- \to \nu_\tau
 \,\mathrm{hadrons}] \over \Gamma [\tau^- \to \nu_\tau e^-
 {\bar \nu}_e] } \, =\,
 R_{\tau,V} + R_{\tau,A} + R_{\tau,S}\, .
$$ 
The theoretical analysis involves the two-point correlation functions for
the vector $\, V^{\mu}_{ij} = \bar{\psi}_j \gamma^{\mu} \psi_i \, $
and axial-vector
$\, A^{\mu}_{ij} = \bar{\psi}_j \gamma^{\mu} \gamma_5 \psi_i \,$
colour-singlet quark currents ($i,j=u,d,s$):
\be\label{eq:pi_v}
\Pi^{\mu \nu}_{ij,\cJ}(q) \equiv
 i \int d^4x \, e^{iqx}
\langle 0|T(\cJ^{\mu}_{ij}(x) \cJ^{\nu}_{ij}(0)^\dagger)|0\rangle  ,
\ee
which have the Lorentz decompositions
\beqn\label{eq:lorentz}
\Pi^{\mu \nu}_{ij,\cJ}(q) & \!\!\!\! = & \!\!\!\!
  (-g^{\mu\nu} q^2 + q^{\mu} q^{\nu}) \, \Pi_{ij,\cJ}^{(1)}(q^2)   \no\\
  && \!\!\!\! +   q^{\mu} q^{\nu} \, \Pi_{ij,\cJ}^{(0)}(q^2) ,
\eeqn
where the superscript $(J=0,1)$ denotes the angular momentum in the hadronic rest frame.

The imaginary parts of 
$\, \Pi^{(J)}_{ij,\cJ}(q^2) \, $
are proportional to the spectral functions for hadrons with the corresponding
quantum numbers.  The semihadronic decay rate of the $\tau$
can be written as an integral of these spectral functions
over the invariant mass $s$ of the final-state hadrons:
\beqn\label{eq:spectral}
R_\tau  &\!\!\!\!\! = &\!\!\!\!\!
12 \pi \int^{m_\tau^2}_0 {ds \over m_\tau^2 } \,
 \left(1-{s \over m_\tau^2}\right)^2
\no\\ &\!\!\!\!\! \times &\!\!\!\!\!
\biggl[ \left(1 + 2 {s \over m_\tau^2}\right)
 \mbox{\rm Im} \Pi^{(1)}(s)
 + \mbox{\rm Im} \Pi^{(0)}(s) \biggr]  .
\eeqn
 The appropriate combinations of correlators are
\beqn\label{eq:pi}
\Pi^{(J)}(s)  &\!\!\! \equiv  &\!\!\!
  |V_{ud}|^2 \, \left( \Pi^{(J)}_{ud,V}(s) + \Pi^{(J)}_{ud,A}(s) \right)
\no\\ &\!\!\! + &\!\!\!
|V_{us}|^2 \, \left( \Pi^{(J)}_{us,V}(s) + \Pi^{(J)}_{us,A}(s) \right).
\eeqn
 The contributions coming from the first two terms correspond to
$R_{\tau,V}$ and $R_{\tau,A}$ respectively, while
$R_{\tau,S}$ contains the remaining Cabibbo-suppressed contributions.

The integrand in Eq.~(\ref{eq:spectral}) cannot be calculated at present from QCD.
Nevertheless the integral itself can be calculated systematically by exploiting
the analytic properties of the correlators $\Pi^{(J)}(s)$. They are analytic
functions of $s$ except along the positive real $s$-axis, where their
imaginary parts have discontinuities.
$R_\tau$ can then be written as a contour integral
in the complex $s$-plane running
counter-clockwise around the circle $|s|=m_\tau^2$:
\beqn\label{eq:circle}
 R_\tau &\!\!\!\!\! =&\!\!\!\!\!
6 \pi i \oint_{|s|=m_\tau^2} {ds \over m_\tau^2} \,
 \left(1 - {s \over m_\tau^2}\right)^2
\no \\ &\!\!\!\!\!\times &\!\!\!\!\!
\left[ \left(1 + 2 {s \over m_\tau^2}\right) \Pi^{(0+1)}(s)
         - 2 {s \over m_\tau^2} \Pi^{(0)}(s) \right] \! .
\eeqn
This expression requires the correlators only for
complex $s$ of order $m_\tau^2$, which is significantly larger than the scale
associated with non-perturbative effects.
Using the Operator Product Expansion (OPE),
$\Pi^{(J)}(s) = \sum_{D} C_D^{(J)}/ (-s)^{D/2}$,
to evaluate the contour integral, $R_\tau$
can be expressed as an expansion in powers of $1/m_\tau^2$.
%
The uncertainties associated with the use of the OPE near the
time-like axis are heavily suppressed by the presence in (\ref{eq:circle})
of a double zero at $s=m_\tau^2$.

The combination $R_{\tau,V+A}$
can be written as \cite{BNP:92}
\begin{equation}\label{eq:Rv+a}
 R_{\tau,V+A} \, =\, N_C\, |V_{ud}|^2\, S_{\mathrm{EW}} \left\{ 1 +
 \delta_{\mathrm{P}} + \delta_{\mathrm{NP}} \right\} ,
\end{equation}
where $N_C=3$ is the number of quark colours
and $S_{\mathrm{EW}}=1.0201\pm 0.0003$ contains the
electroweak radiative corrections \cite{MS:88,BL:90,ER:02}.
The dominant correction ($\sim 20\%$) is the perturbative QCD
contribution $\delta_{\mathrm{P}}$, which is already known to
$O(\alpha_s^4)$ \cite{BNP:92,BChK:08} and includes a resummation of the most
important higher-order effects \cite{LDP:92a,PI:92}.

Non-perturbative contributions are suppressed by six powers of the
$\tau$ mass \cite{BNP:92} and, therefore, are very small. Their
numerical size has been determined from the invariant-mass
distribution of the final hadrons in $\tau$ decay, through the study
of weighted integrals \cite{LDP:92b},
\begin{equation}\label{eq:moments}
 R_{\tau}^{kl} \,\equiv\, \int_0^{m_\tau^2} ds\, \left(1 - {s\over
 m_\tau^2}\right)^k\, \left({s\over m_\tau^2}\right)^l\, {d
 R_{\tau}\over ds} \, ,
\end{equation}
which can be calculated theoretically in the same way as $R_{\tau}$.
The predicted suppression \cite{BNP:92} of the non-perturbative
corrections has been confirmed by ALEPH \cite{ALEPH:05}, CLEO
\cite{CLEO:95} and OPAL \cite{OPAL:98}. The most recent analysis
\cite{DHZ:05} gives
\begin{equation}\label{eq:del_np}
 \delta_{\mathrm{NP}} \, =\, -0.0059\pm 0.0014 \, .
\end{equation}
%

%
\begin{figure}[tb]
\label{fig:alpha_s} \centering
\includegraphics[width=7.6cm]{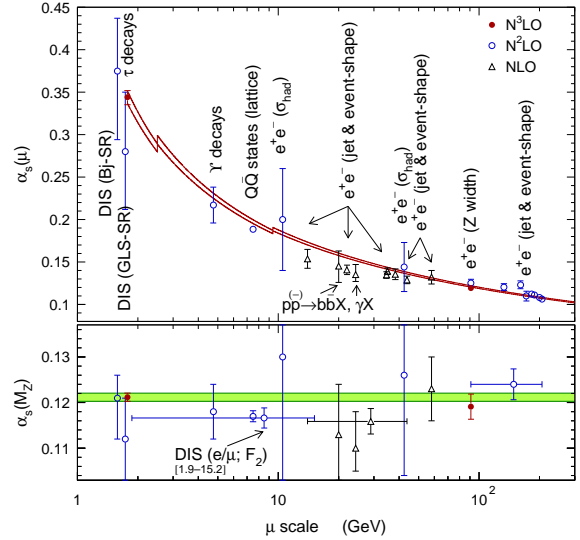}
\vspace{-0.9cm}
\caption{Measured values of $\alpha_s$ at different
scales. The curves show the energy dependence predicted by QCD,
using $\alpha_s(m_\tau^2)$ as input. The corresponding extrapolated
$\alpha_s(M_Z^2)$ values are shown at the bottom, where the shaded
band displays the $\tau$ decay result within errors \cite{DHZ:05}.}
\end{figure}

The QCD prediction for $R_{\tau,V+A}$ is then completely dominated
by $\delta_P$; 
non-perturbative effects being
smaller than the perturbative uncertainties from uncalculated
higher-order corrections. The result turns out to be very sensitive
to the value of $\alpha_s(m_\tau^2)$, allowing for an accurate
determination of the fundamental QCD coupling \cite{NP:88,BNP:92}.
The experimental measurement $R_{\tau,V+A}= 3.479\pm0.011$ implies
\cite{DHZ:05}
\begin{equation}\label{eq:alpha}
 \alpha_s(m_\tau^2)  \,=\,  0.344\pm
 0.005_{\mathrm{exp}}\pm 0.007_{\mathrm{th}} \, .
\end{equation}

The strong coupling measured at the $\tau$ mass scale is
significantly larger than the values obtained at higher energies.
From the hadronic decays of the $Z$, one gets $\alpha_s(M_Z^2) =
0.1191\pm 0.0027$ \cite{BChK:08,DHZ:05,LEPEWWG}, which differs from $\alpha_s(m_\tau^2)$
by more than $20\,\sigma$. 
After evolution up to the scale $M_Z$ \cite{Rodrigo:1998zd}, the strong
coupling constant in (\ref{eq:alpha}) decreases to \cite{DHZ:05}
\begin{equation}\label{eq:alpha_z}
 \alpha_s(M_Z^2)  \, =\,  0.1212\pm 0.0011 \, ,
\end{equation}
in excellent agreement with the direct measurements at the $Z$ peak
and with a better accuracy. The comparison of these two
determinations of $\alpha_s$ in two very different energy regimes, $m_\tau$
and $M_Z$, provides a beautiful test of the predicted running of the
QCD coupling; i.e., a very significant experimental verification of
{\it asymptotic freedom}.

\section{Perturbative contribution to $R_\tau$}

The recent calculation of the $\cO(\alpha_s^4)$ contribution to $\Pi^{(0+1)}(s)$
\cite{BChK:08} has triggered a renewed theoretical interest on $R_\tau$ \cite{BChK:08,DHZ:05,BJ:08,MY:08}.
The perturbative contribution $\delta_P$ is extracted from the Adler function
\be\label{eq:d}
- s {d \over ds } \Pi^{(0+1)}(s)
=  {1\over 4 \pi^2} \sum_{n=0}  K_n
\left( {\alpha_s(s)\over \pi}\right)^n  .
\ee
%
For three flavours, the known coefficients take the values:
$K_0 = K_1 = 1$; $K_2 = 1.63982$; 
$K_3(\overline{MS}) = 6.37101$ and $K_4(\overline{MS}) =49.07570$  \cite{BChK:08}.

The perturbative component of $R_\tau$ is given by
\be\label{eq:r_k_exp}
\delta_P =
\sum_{n=1}  K_n \, A^{(n)}(\alpha_s) ,
\ee
where the functions \cite{LDP:92a}
\beqn\label{eq:a_xi}
A^{(n)}(\alpha_s) &\!\!\!\! = &\!\!\!\! {1\over 2 \pi i}
\oint_{|s| = m_\tau^2} {ds \over s} \,
  \left({\alpha_s(-s)\over\pi}\right)^n
\no\\ &\!\!\!\!\times  &\!\!\!\!
 \left( 1 - 2 {s \over m_\tau^2} + 2 {s^3 \over m_\tau^6}
         - {s^4 \over  m_\tau^8} \right)
\eeqn
are contour integrals in the complex plane, which only depend on
$a_\tau\equiv\alpha_s(m_\tau^2)/\pi$. Using the exact solution
(up to unknown $\beta_{n>4}$ contributions) for $\alpha_s(s)$
given by the renormalization-group $\beta$-function equation,
they can be numerically computed with a very high accuracy \cite{LDP:92a}.
One can easily check that the results are very stable under changes of the renormalization scale
and rather insensitive to the truncation of the $\beta$ function (putting $\beta_4=0$
has a negligible impact). Thus, the resulting theoretical uncertainty on $\delta_P$ is small.

However if, instead of adopting the known values for $A^{(n)}(\alpha_s)$, one expands $\alpha_s(-s)$  in powers of
$\alpha_s(m_\tau)$ inside the  the integrals (\ref{eq:a_xi}), the large logarithmic running along the circle $s=m_\tau^2\exp{(i\phi)}$
($\phi\epsilon [0, 2\pi]$) gives rise to a nearly divergent series of the form
$\delta_P \, = \,\sum_{n=1} (K_n + g_n) \, a_\tau^n$,
where the $g_n$ coefficients depend on $K_{m<n}$ and on $\beta_{m<n}$:
\be\label{eq:delta_0}
\delta^{(0)} = a_\tau + 5.20 \, a_\tau^2 + 26.4 \, a_\tau^3
 + 127 \, a_\tau^4  + \, \cdots
\ee
The ``running'' $g_n$ contributions are much larger than the original $K_n$ coefficients
containing the Adler function dynamics ($g_2 = 3.563$, $g_3 = 19.99$, $g_4=78.00$) \cite{LDP:92a}.
These generates a sizeable renormalization scale dependence, which is much larger
than the naively expected $\cO(\alpha_s^5)$ effect.
The radius of convergence of this expansion is actually quite small.
A numerical analysis of the series \cite{LDP:92a}
shows that, at the three-loop level, an upper estimate for the convergence radius  is
$a_{\tau,\mbox{\rms conv}} \, < 0.11$, which is very close to the physical value.
Thus, the fixed-order expansion \eqn{eq:delta_0} should not be used for accurate predictions of $R_\tau$.
The result \eqn{eq:alpha} has been correctly obtained using Eq.~\eqn{eq:r_k_exp} with the exact values of
the functions $A^{(n)}(\alpha_s)$. The slightly different results quoted in refs.~\cite{BChK:08,BJ:08}
originate in their use of the pathological fixed-order expansion \eqn{eq:delta_0}.\footnote{
A better convergence of the fixed-order expansion \eqn{eq:delta_0} is enforced in Ref.~\cite{BJ:08} through an artificial cancelation of the $K_n$ and $g_n$ contributions at higher orders. Since $R_\tau$ does not get corrections from $D=4$ terms in the OPE, this behaviour is trivially accomplished assuming that the perturbative series is dominated by an $n=2$ IR renormalon. While this provides an interesting academic model of higher-order contributions, the resulting wild behaviour of the Adler series is totally ad-hoc and generates problems for weighted distributions of the form \eqn{eq:moments}. The non-perturbative correction in \eqn{eq:del_np} would no longer be valid within this model, making the low value of $\alpha_s(m_\tau)$ claimed in \cite{BJ:08} unjustified.}

\section{$|V_{us}|$ determination from 
tau decays}

The separate measurement of the $|\Delta S|=0$ and $|\Delta S|=1$
tau decay widths provides a very clean determination of $V_{us}\,$
\cite{GJPPS:05,PI:07b}.
To a first approximation the Cabibbo mixing can be directly obtained
from experimental measurements, without any theoretical input.
Neglecting the small SU(3)-breaking corrections from the $m_s-m_d$
quark-mass difference, one gets:
$$ 
 |V_{us}|^{\mathrm{SU(3)}} =\: |V_{ud}| \:\left(\frac{R_{\tau,S}}{R_{\tau,V+A}}\right)^{1/2}
 =\: 0.210\pm 0.003\, .
$$ 
We have used $|V_{ud}| = 0.97418\pm 0.00027$ \cite{PDG},    
$R_\tau = 3.640\pm 0.010$       
and the value $R_{\tau,S}=0.1617\pm 0.0040$ \cite{PI:07b}, which results from the
most recent BaBar \cite{BA:07} and Belle \cite{BE:07} measurements of Cabibbo-suppressed
tau decays \cite{Banerjee}.
The new branching ratios measured by BaBar and Belle are all smaller than the previous
world averages, which translates into a smaller value of $R_{\tau,S}$ and $|V_{us}|$.
For comparison, the previous value $R_{\tau,S}=0.1686\pm 0.0047$ \cite{DHZ:05} resulted in $|V_{us}|^{\mathrm{SU(3)}}=0.215\pm 0.003$.

This rather remarkable determination is only slightly shifted by
the small SU(3)-breaking contributions induced by the strange quark mass.
These corrections can be 
estimated through a QCD analysis of the differences
\cite{GJPPS:05,PI:07b,PP:99,ChDGHPP:01,ChKP:98,KKP:01,MW:06,KM:00,MA:98,BChK:05}
\begin{equation}
 \delta R_\tau^{kl}  \,\equiv\,
 {R_{\tau,V+A}^{kl}\over |V_{ud}|^2} - {R_{\tau,S}^{kl}\over |V_{us}|^2}\, .
\end{equation}
%
The only non-zero contributions are proportional 
to the mass-squared difference $m_s^2-m_d^2$ or to vacuum expectation
values of SU(3)-breaking operators such as $\delta O_4
\equiv \langle 0|m_s\bar s s - m_d\bar d d|0\rangle \approx (-1.4\pm 0.4)
\cdot 10^{-3}\; \mathrm{GeV}^4$ \cite{PP:99,GJPPS:05}. The dimensions of these operators
are compensated by corresponding powers of $m_\tau^2$, which implies a strong
suppression of $\delta R_\tau^{kl}$ \cite{PP:99}:
\beqn\label{eq:dRtau}
 \delta R_\tau^{kl} &\!\!\approx &\!\!  24\, S_{\mathrm{EW}}\; \left\{ {m_s^2(m_\tau^2)\over m_\tau^2} \,
 \left( 1-\epsilon_d^2\right)\,\Delta_{kl}(\alpha_s)
 \right.\no\\ &&\hskip 1.3cm\left.
 - 2\pi^2\, {\delta O_4\over m_\tau^4} \, Q_{kl}(\alpha_s)\right\}\, ,
\eeqn
where $\epsilon_d\equiv m_d/m_s = 0.053\pm 0.002$ \cite{LE:96}.
The perturbative 
corrections $\Delta_{kl}(\alpha_s)$ and
$Q_{kl}(\alpha_s)$ are known to $O(\alpha_s^3)$ and $O(\alpha_s^2)$,
respectively \cite{PP:99,BChK:05}.

The $J=0$ contribution to $\Delta_{00}(\alpha_s)$ shows a rather
pathological behaviour, with clear signs of being a non-convergent perturbative
series. Fortunately, the corresponding longitudinal contribution to
$\delta R_\tau\equiv\delta R_\tau^{00}$ can be estimated phenomenologically with a much better
accuracy, $\delta R_\tau|^{L}\, =\, 0.1544\pm 0.0037$ \cite{GJPPS:05,JOP:06},
because it is dominated by far by the well-known $\tau\to\nu_\tau\pi$
and $\tau\to\nu_\tau K$ contributions. To estimate the remaining transverse
component, one needs an input value for the strange quark mass. Taking the
range
$m_s(m_\tau) = (100\pm 10)\:\mathrm{MeV}$ \
[$m_s(2\:\mathrm{GeV}) = (96\pm 10)\:\mathrm{MeV}$],
which includes the most recent determinations of $m_s$ from QCD sum rules
and lattice QCD \cite{JOP:06},
one gets finally $\delta R_{\tau,th} = 0.216\pm 0.016$, which implies \cite{PI:07b}
\beqn\label{eq:Vus_det}
 |V_{us}| &=& \left(\frac{R_{\tau,S}}{\frac{R_{\tau,V+A}}{|V_{ud}|^2}-\delta
 R_{\tau,\mathrm{th}}}\right)^{1/2}
 \no\\ &=&
 0.2165\pm 0.0026_{\mathrm{\, exp}}\pm 0.0005_{\mathrm{\, th}}\, .
\eeqn
%
A larger central value,
$|V_{us}| = 0.2212\pm 0.0031$, 
is obtained with the old world average for $R_{\tau,S}$.

Sizeable changes on the experimental determination of $R_{\tau,S}$ are to be expected from
the full analysis of  the huge BaBar and Belle data samples. In particular, the high-multiplicity
decay modes are not well known at present.
Thus, the result (\ref{eq:Vus_det}) could easily fluctuate in the near future.
However, it is important to realize that the final error of the $V_{us}$ determination from
$\tau$ decay is completely dominated by the experimental uncertainties. If $R_{\tau,S}$
is measured with a 1\% precision, the resulting $V_{us}$ uncertainty will
get reduced to around 0.6\%, i.e. $\pm 0.0013$, making $\tau$ decay the best source of
information about $V_{us}$.

An accurate measurement of the invariant-mass distribution of the final hadrons
could make possible a simultaneous determination
of $V_{us}$ and the strange quark mass, through a correlated analysis of
several weighted differences $\delta R_\tau^{kl}$. The extraction of $m_s$ suffers from
theoretical uncertainties related to the convergence of the perturbative series
$\Delta_{kl}(\alpha_s)$, which makes necessary a better
understanding of these corrections.

\section{$\tau\to\nu_\tau K\pi$ and $K\to\pi l\bar\nu_l$}

The decays $\tau\to\nu_\tau K\pi$ probe the same hadronic form factors investigated in
$K_{l3}$ processes, but they are sensitive to a much broader range of invariant masses.
A theoretical understanding of the form factors can be achieved, using analyticity, unitarity and some general properties of QCD, such as chiral symmetry and the short-distance asymptotic behaviour \cite{taurev06,taurev98}.

\begin{figure}[tbh]
\centering
\includegraphics[width=7.6cm]{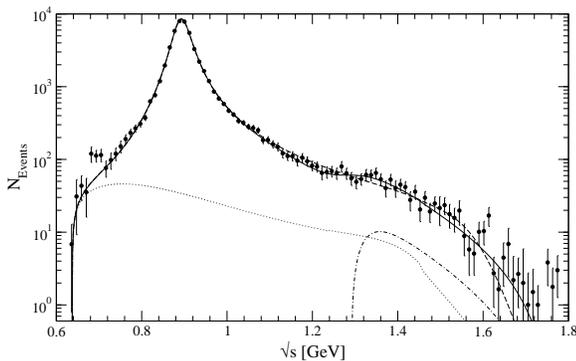}
\vspace{-0.9cm}
\caption{Theoretical description \cite{JPP:08} (solid line) of the Belle
$\tau^-\to\nu_\tau K_S\pi^-$ data \cite{BE:07}. The $K^{*'}$ (dashed-dotted) and scalar (dotted) contributions are also shown. \label{fig:KpSpectrum}}
\end{figure}

Figure~\ref{fig:KpSpectrum} compares the resulting theoretical description of the $\tau$ decay spectrum \cite{JPP:08}  with the recent Belle measurement \cite{BE:07}. At low values of $s$ there is clear evidence of the scalar contribution, which was predicted previously using a careful analysis of $K\pi$ scattering data \cite{JOP:06,JOP:00}. From the measured $\tau$ spectrum one obtains $M_{K^*} = 895.3\pm 0.2$ MeV and $\Gamma_{K^*}=47.5\pm 0.4$ MeV  \cite{JPP:08}. Since the absolute normalization is fixed by
$K_{l3}$ data to be $|V_{us}|\, f_+^{K^0\pi^-}(0)= 0.21664 \pm 0.00048$ \cite{Kl3}, one gets then a theoretical prediction for the branching fraction,
Br$(\tau^-\to\nu_\tau K_S\pi^-) = 0.427 \pm 0.024\%$, in good agreement with the Belle measurement $0.404 \pm 0.013\%$, although slightly larger.

The $\tau$ determination of the vector form factor $f_+^{K\pi}(s)$ \cite{JPP:08,BEJ:08}
provides precise values for its slope and curvature,  $\lambda_+'= (25.2 \pm 0.3)\cdot 10^{-3}$ and $\lambda_+''=
 (12.9 \pm 0.3)\cdot 10^{-4}$  \cite{JPP:08}, in agreement but more precise than the corresponding $K_{l3}$ measurements~\cite{Kl3}.

%


\section*{Acknowledgements}
This work has been supported
by MICINN, Spain (grants FPA2007-60323 and
Consolider-Ingenio 2010 CSD2007-00042, CPAN) by the
EU Contract MRTN-CT-2006-035482 (FLAVIAnet) and by Generalitat Valenciana
(PROMETEO/2008/069).



\begin{thebibliography}{9}

\bibitem{PhiPsi08} A. Pich,
\NPPS\ 181-182 (2008) 300;
169 (2007) 393.

\bibitem{taurev06} A. Pich, 
   \MP\ A 21 (2006) 5652.

\bibitem{taurev98} A. Pich, {\it Tau Physics}, in {\it Heavy Flavours II},
eds. A.J.~Buras and M.~Lindner,
Advanced Series on Directions in High Energy Physics  15   
(World Scientific, Singapore, 1998) p.~453, arXiv:hep-ph/9704453.


\bibitem{BR:88} E. Braaten, Phys. Rev. Lett. 60 (1988) 1606;
           Phys. Rev. D 39 (1989) 1458.

\bibitem{NP:88} S. Narison and A. Pich, Phys. Lett. B 211 (1988) 183.

\bibitem{BNP:92}
 E. Braaten, S. Narison and A. Pich, Nucl. Phys. B 373 (1992) 581.

\bibitem{LDP:92a} F. Le Diberder and A. Pich, Phys. Lett. B 286 (1992) 147.

\bibitem{QCD:94} A. Pich, Nucl. Phys. B (Proc. Suppl.) 39B,C (1995) 326.

\bibitem{MS:88} W.J. Marciano and A. Sirlin, Phys. Rev. Lett. 61 (1988) 1815.

\bibitem{BL:90} E. Braaten and C.S. Li, Phys. Rev. D 42 (1990) 3888.

\bibitem{ER:02} J. Erler, Rev. Mex. Phys. 50 (2004) 200. 

\bibitem{BChK:08} P.A. Baikov, K.G. Chetyrkin and J.H. K\"uhn, 
 Phys. Rev. Lett. 101 (2008) 012002.

\bibitem{PI:92} A.A. Pivovarov, Z. Phys. C 53 (1992) 461.

\bibitem{LDP:92b} F. Le Diberder and A. Pich, Phys. Lett. B 289 (1992) 165.

\bibitem{ALEPH:05}
 ALEPH Collaboration, Phys. Rep. 421 (2005) 191; 
 Eur. Phys. J. C 4 (1998) 409; Phys. Lett. B 307 (1993) 209.

\bibitem{CLEO:95} CLEO Collaboration, Phys. Lett. B 356 (1995) 580.

\bibitem{OPAL:98} OPAL Collaboration, Eur. Phys. J. C 7 (1999) 571. 

\bibitem{DHZ:05} M. Davier et al.,
Rev. Mod. Phys. 78 (2006) 1043; 
Eur. Phys. J. C 56 (2008) 305. 

\bibitem{LEPEWWG}
The LEP Collaborations ALEPH, DELPHI, L3 and OPAL and the LEP
Electroweak Working Group, arXiv:0712.0929 [hep-ex];\\
http://www.cern.ch/LEPEWWG/.

\bibitem{Rodrigo:1998zd}
  G.~Rodrigo, A.~Pich and A.~Santamaria, Phys. Lett. B 424 (1998) 367.

\bibitem{BJ:08}
M. Beneke and M. Jamin, JHEP 0809 (2008) 044.
\bibitem{MY:08}
K.~Maltman and T. Yavin, arXiv:0807.0650 [hep-ph].

\bibitem{GJPPS:05}
E. G\'amiz et al., Phys. Rev. Lett. 94 (2005) 011803; 
 JHEP 0301 (2003) 060. 

\bibitem{PI:07b} E. G\'amiz et al., PoS KAON 008 (2007).

\bibitem{PDG}
C. Amsler et al.,  {\em The Review of Particle Physics}, Phys. Lett. B 667, 1 (2008).

\bibitem{BA:07} BaBar Collaboration,
 Phys. Rev. Lett. 100 (2008) 011801;  
 Phys. Rev. D 76 (2007) 051104. 

\bibitem{BE:07} Belle Collaboration,
 Phys. Lett. B 654 (2007) 65; 
 643 (2006) 5. 

\bibitem{Banerjee} S. Banerjee, PoS KAON 009 (2007). 


\bibitem{PP:99} A. Pich and J. Prades,
 JHEP 9910 (1999) 004; 
    9806 (1998) 013. 

\bibitem{ChDGHPP:01} S. Chen et al., Eur. Phys. J. C 22 (2001) 31.
M. Davier et al., Nucl. Phys. B (Proc. Suppl.) 98 (2001) 319.

\bibitem{ChKP:98} K.G. Chetyrkin, J.H. K\"uhn and A.A.~Pivovarov,
  Nucl. Phys. B 533 (1998) 473.  

\bibitem{KKP:01} J.G.~K\"orner, F. Krajewski and A.A. Pivovarov,
  Eur. Phys. J. C 20 (2001) 259. 

\bibitem{MW:06} K. Maltman and C.E. Wolfe, Phys. Lett. B 639 (2006) 283.
  K. Maltman et al. arXiV:0807.3195 [hep-ph].

\bibitem{KM:00} J. Kambor and K. Maltman,
   Phys. Rev. D 62 (2000) 093023; 
   64 (2001) 093014. 

\bibitem{MA:98}
K. Maltman, Phys. Rev. D 58 (1998) 093015. 

\bibitem{BChK:05}
P.A. Baikov, K.G. Chetyrkin and J.H.~K\"uhn, Phys. Rev. Lett. 95 (2005)
012003. 

\bibitem{LE:96} H. Leutwyler, Phys. Lett. B 378 (1996) 313. 

\bibitem{JOP:06} M. Jamin, J.A. Oller and A. Pich, Phys. Rev. D 74 (2006) 074009.


\bibitem{JPP:08} M. Jamin, A. Pich and J. Portol\'es,
   Phys. Lett. B 640 (2006) 176;
   664 (2008) 78.  


\bibitem{JOP:00} M. Jamin, J.A. Oller, and A. Pich,
Nucl. Phys. B 587 (2000) 331; 
622 (2002) 279; 
Eur. Phys. J. C 24 (2002) 237. 

\bibitem{Kl3}
FlaviaNet Working Group on Kaon Decays, \NPPS\ 181--182 (2008) 83.

\bibitem{BEJ:08} D.R. Boito et al. arXiv:0807.4883 [hep-ph].




\end{thebibliography}
\end{document}